\documentclass[a4paper,12pt]{article}
\usepackage{epsfig,amsmath,amssymb,subfigure,placeins}
\usepackage{graphicx}
\usepackage{dcolumn}
\usepackage{bm}
\usepackage{subfig}
\usepackage{float}
\usepackage{tabularx}
\usepackage{multirow}
\usepackage{hhline}
\usepackage{color}
\newcommand{\eps}{\epsilon_{s}}

\newcommand{\old}[1]{}

\newcommand{\be}{\begin{equation}}
\newcommand{\ee}{\end{equation}}
\newcommand{\ba}{\begin{eqnarray}}
\newcommand{\ea}{\end{eqnarray}}
\newcommand{\bi}{\begin{itemize}}
\newcommand{\ei}{\end{itemize}}

\newcommand{\pt}{p_{{}_T}}
\newcommand{\ptmin}{p_{{}_T}^{\mbox{min}}}
\newcommand{\ptmax}{p_{{}_T}^{\mbox{max}}}
\newcommand{\spt}{S(p_{{}_T})}
\newcommand{\sinpt}{ \langle S\rangle}
\newcommand{\sincl}{{\langle S \rangle}^{{}^{\mbox{incl}}} }

\begin{document}
\begin{flushright}
{\normalsize
}
\end{flushright}
\vskip 0.1in
\begin{center}
{\large{\bf  Bottomonium suppression: A probe to the pre-equilibrium 
era of quark matter}}
\end{center}
\vskip 0.1in
\begin{center}
Uttam Kakade\footnote{usk11dph@iitr.ac.in}
Binoy Krishna Patra\footnote{binoyfph@iitr.ac.in}
and Lata Thakur\footnote{lata1dph@iitr.ac.in}\\
{\small {\it Department of Physics, Indian Institute of
Technology Roorkee, India, 247 667} }
\end{center}
\vskip 0.01in

\addtolength{\baselineskip}{0.4\baselineskip} 
\section* {Abstract}
We have studied the thermal suppression of the bottomonium 
states in relativistic heavy-ion collision at LHC energies
as function of centrality, rapidity, transverse momentum etc.
to explain the CMS data. Our investigation mainly 
spans over three problems: a) how the theoretical 
predictions might still be modified by the remnants of the non-perturbative
confining force, b) how does the presence of a not necessarily isotropic 
QCD medium modify the potential (both the real and imaginary part) acting 
between a static quark and antiquark pair, and c) finally how the additional 
time-zone of pre-equilibrium partonic evolution, in addition to the above 
modifications, affects the bottomonium production at the LHC energies.
We resolve them by correcting both the perturbative 
and nonperturbative terms of the $Q\bar Q$ potential in 
(an)isotropic QCD medium and then couple to the dynamics of
the system undergoing successive pre-equilibrium 
and equilibrium era. Due to the tiny 
formation time and the large binding energy of bottomonium (1S) state,
we succeed in constraining the isotropization time and the shear 
viscosity.\\

PACS:~~ 12.39.-x,11.10.St,12.38.Mh,12.39.Pn
12.39.Hg; 12.38.Gc \\

\section{Introduction}
The experimental programs at Super Proton Synchrotron (SPS) at CERN, 
Relativistic Heavy Ion Collider (RHIC) at BNL and Large Hadron 
Collider (LHC) at CERN open up an window onto the properties of 
Quantum Chromodynamics (QCD) at high temperatures in guise of 
quark-gluon plasma (QGP). Following the conjecture of Matsui and Satz
~\cite{Matsui:86}, there was considerable interest to study
the properties of quarkonia at finite temperature. 
It was until recently that the inherent hierarchy of the scales 
in the heavy quark bound systems ($m_Q \gg m_Q v \gg m_Q v^2$) 
facilitates to derive a sequence of 
effective field theories (EFT) from the underlying theory, QCD, {\em namely}
non relativistic QCD (NRQCD) and potential NRQCD (pNRQCD), by
integrating out the successive scales in the system.
The heavy quark bound states are described by the singlet 
and octet potentials through the matching coefficients 
in the effective lagrangian, which, however can be extended to 
finite temperature~\cite{Brambilla:2008cx}
with the additional thermal scales, $T$, $gT$, $g^2T$ etc.
The thermal corrections to the real
and imaginary part of the singlet potential are manifested 
as the Debye screening~\cite{Matsui:86} and the Landau damping~
\cite{Beraudo:2007ky,Laine:2007qy}, respectively.
On the other hand, the non EFT defines the potential 
from the late time behavior of a Wilson
loop~\cite{Brambilla:2008cx,Wilson:1974sk,Makeenko,Berges,Barchielli:NPB296,Rothkopf:MPL28,Laine:2006ns}. 
However, at finite temperature, the Wilson loop depends on imaginary time and
the analytic continuation in the large real-time limit gives 
the (complex) potential~\cite{Brambilla:2008cx,Laine:2006ns},
whose imaginary part is manifested as Landau damping \cite{Beraudo:2007ky}.

The separation of thermal scales in EFT 
is not evident and one needs lattice techniques to test 
the approach, where the dissociation of the quarkonium 
states can be studied even without the potential models
rather the physics of a given quarkonium state is encoded 
in its spectral function in terms of the Euclidean 
meson correlation functions~\cite{Karsch,Mocsy05-08,Wong05,Cabrera07,
Alberico08,Forcrand}.
However, the reconstruction of the spectral functions from the lattice meson 
correlators turns out to be very difficult. At finite temperature the 
situation becomes worse because the temporal extent is decreasing and 
particularly, for the bottomonium states, it becomes worst, thus 
inadvertently supports the use of potential models at finite
temperature to complement the lattice studies.

The physical picture of quarkonium dissociation 
has been evolved over the years, where the properties of thermally 
produced heavy quarkonium 
states can be observed through the energy spectrum of their decay products
~\cite{Schenke,Martinez}. Thus the disappearance of the peak
in the resonance peak hints the dissolution of the state.
Physically a resonance is dissolved into
a medium through the broadening of its width. In EFT framework, 
when the binding energy is large compared to the temperature, the 
resonances acquire a finite width 
due to interactions with ultra-soft gluons, causing the singlet-to-octet 
transitions~\cite{Brambilla:2008cx}. This picture is relevant 
for the $\Upsilon$(1S) suppression at the LHC.
But when the binding energy is smaller than any of the 
above thermal scales, the potential acquires an imaginary 
component~\cite{Brambilla:2008cx} (Landau Damping), which induces 
a thermal width. However beyond the leading-order the above
mentioned processes become entangled.
On the other hand, in a non-EFT framework, the width arises either 
when a bound state absorbs a hard gluons or a light parton of the medium
scatters off the bound state by exchanging a space-like gluon.

The (heavy) quark and antiquark ($Q \bar Q$) pairs are produced in 
heavy ion collisions on a very short time-scale ($\sim 1/2m_{Q}$),
when the initial state effects on the parton densities 
(shadowing)~\cite{vogt:PRC812010}, the initial state 
energy loss~\cite{vogt:PRC612000}, the intrinsic heavy flavors, and the final 
state absorption on nucleons~\cite{vogt:PRC812010,sgavin:PRL681992} etc.
could affect the production mechanism intimately.
The shadowing and absorption are important at mid rapidity
whereas the (initial-state) energy loss and intrinsic heavy
flavor are important at forward rapidity.
As the times are elapsed, the resonances are fomed over a 
formation time, $\tau_F$ and traverses the plasma and then the 
hadronic matter before leaving the interacting system to be decayed
into  a dilepton. This long `trek' inside the interacting system is 
cliffhanger for the pair. By the time the resonance 
is formed, either the screening of the color force~\cite{Matsui:86}
or an energetic gluon~\cite{xu,gdiss1-3}, even a comoving hadron~\cite{vogt}
could dissociate the resonance(s). 
Since the expansion of the matter produced in heavy-ion collisions
proceeds through the successive stages of pre equilibrium (anisotropic)
and equilibrium (isotropic) phases, therefore a study of 
quarkonium production is poised to provide a wealth
of information about the evolution of the plasma and its 
in-medium properties.

In the early days of collider experiments at SPS and RHIC, most of the 
interests were focused on the suppression of $c \bar c$ bound 
states \cite{Matsui:86,Karsch:1987pv} but several 
observations are yet to be understood {\em namely}
the suppression of $\psi$ (1S) does not increase from SPS to RHIC,
even though the centre-of-mass energy is increased by fifteen times. The 
heavy-ion program at the LHC 
may resolve those puzzles because the beam energy and luminosity are 
increased by ten times of that
of the RHIC. Moreover the CMS detector has excellent capabilities
for muon detection and provides measurements of $\psi$(2S) and the 
$\Upsilon$ family, which enables the quantitative analysis of quarkonia.
That is why the interest may be shifted to the bottomonium
states at the LHC energy due to the following reasons:
i) The initial state effects to the bottomonium production are much 
smaller than the charmonium production. ii) The 
bottomonium is much heavier than the charmonium, the competition
due to the recombination is thus unlikely.
iii) Since the bottom quark is heavier, it
can be dealt efficiently by the potential approach.
iv) Although the $\Upsilon$ states have diverse
binding energies but their similar decay 
kinematics and production mechanisms 
enable to measure their relative suppression unambiguously.

The works described above were limited to an isotropic medium but 
the system produced in relativistic heavy-ion collision 
may not be homogeneous and isotropic because at the early stages of the 
collision, the asymptotic weak-coupling enhances the longitudinal expansion 
substantially than the radial expansion. There have been significant 
advances in the dynamical models 
used to simulate plasma evolution with the 
momentum-space anisotropies 
in full (3+1)-dimensional simulations
\cite{Martinez:2010sd-12tu,Ryblewski:2010bs,
Ryblewski:2012rr,Florkowski:2010cf}. In recent years,  
the effects of anisotropy on the quarkonia states 
have been extensively investigated~\cite{Dumitru:2007hy-09fy-09ni,
Burnier:2009yu} by the leading-anisotropic 
correction to the perturbative term of the potential alone
and found that the anisotropy can have a significant impact 
on quarkonium suppression.
However, in the experimentally relevant regime of temperature
(just above the crossover or transition temperature), the theoretical 
predictions based on high temperature methods, such as HTL perturbation
theory might be modified by the remnants of the non-perturbative
confining force~\cite{prc_vineet}. 
Although the direct lattice QCD based determinations of the
potential have progressed a lot, a model potential for the phenomenological
descriptions of heavy quarkonium suppression would indeed be
quite useful. This is one of the main goal of this present study
and argue for the modification of the full Cornell potential 
as an appropriate potential for heavy quarkonium at finite temperature.

Recently we have investigated the properties of quarkonia states through 
the medium modifications to both the perturbative and nonperturbative 
terms of the $Q \bar Q$ potential~\cite{lata:arxiv2013} in the presence of a not necessarily isotropic QCD 
medium, which have mainly two important observations: The first one is 
that the inclusion of the confining string term, in addition to the Coulomb
term makes both the real and imaginary parts of the potential more 
stronger, compared to the medium correction of the perturbative term 
of the potential alone~\cite{thakur:PRD2013}.
Since the imaginary part contributes to the width ($\Gamma$) of quarkonium
bound states~\cite{Beraudo:2007ky,Laine:2007qy,Laine:2006ns} which
in turn determines the dissociation temperatures, so
the above cumulative effects due to the remnants of nonperturbative term 
dissociate the quarkonia states at higher temperatures. 
Secondly the presence of the anisotropy makes the 
real-part of the potential stronger but the imaginary-part becomes leaner
and overall the anisotropy makes the quarkonia to dissociate 
at higher temperatures, compared to the isotropic medium.
In the present work, we continue with our model 
potential~\cite{lata:arxiv2013} and numerically obtain
the dissociation temperatures of the ground and the excited states
of the $\Upsilon$ family (which was not done earlier 
in~\cite{lata:arxiv2013}). With these understandings about the quarkonia
states in a static (an)isotropic medium, we move on to study the dynamical 
(sequential) suppression of the bottomonium states in 
nucleus-nucleus collisions at the LHC energies.
We found that the local equilibrium hydrodynamic regime alone 
may not be sufficient to suppress the bottomonium states adequately 
and some additional (pre-equilibrium) time zone of plasma evolution 
needs to be scanned, which seems plausible theoretically
as well as experimentally. The unique features of 
the bottomonium (1S) state, {\em namely} the tiny formation time and 
large binding energy, facilitate to probe both 
the  (pre-equilibrium) era prior to the isotropization time 
and the shear viscosity-to-the entropy ratio.

Our work is organized as follows. In Section~2, we 
revisited the anisotropic corrections to the retarded, advanced and
symmetric gluon self energies and the corresponding (static) propagators
in hard thermal loop perturbation theory~\cite{Carrington:PRD582009} and then study 
the in-medium properties 
of the quarkonium states by the resulting complex potential (in Section~
2.1 and 2.2, respectively). In Section~2.3, we study the dissociation
through a complex potential by obtaining the real and imaginary part of 
the binding energies and
calculate the dissociation temperatures of the ground and excited 
$b \bar b$ states.
Next we switch over our discussion (in Section 3) to an expanding 
medium, which undergoes expansion through the successive pre-equilibrium 
and equilibrium era and
study the survival of the bottomonium states by coupling the in-medium 
dissociation with the dynamics of the expansion.
We found that our model explains the CMS data~\cite{CMS:2012} reasonably 
well, apart from the uncertainties arising due to various initial-state effects,
which is however expected to be very small for the bottomonium states
at the LHC. Finally, we conclude in Section 4.
\section{Bottomonium in anisotropic medium}
An interesting property at the initial phase of the QGP is nowadays 
the anisotropies that occur~\cite{Arnold05,Romatschke;2007mq,rebhan}.
It is therefore worthwhile to consider the properties of quarkonia
such as the binding energy, decay width in such a system. The calculation 
of the real part of the potential 
at finite  anisotropy was first obtained in 
Ref. \cite{Dumitru:2007hy-09fy-09ni,Romatschke;2007mq,laine2} and  was later
extended for the imaginary
part~\cite{Brambilla:2008cx,Dumitru:2007hy-09fy-09ni, Laine09}, by 
the leading anisotropic
corrections to the perturbative term of the potential alone, which 
was further coupled with the dynamical evolution of the anisotropic 
plasma to quantify the quarkonium suppression in nuclear 
collisions~\cite{Strickland:2011mw,Strickland:2011aa}. We now continue 
with the above works to derive the potential at finite temperature 
keeping both the perturbative and nonperturbative terms via the Keyldesh 
presentation in real-time formalism.
\subsection{Real part of the potential}
Since the mass of the heavy quark is very large, so both the 
requirements: $m_Q \gg \Lambda_{QCD}$ and $T \ll m_Q$ are met for the description of the 
interactions between a pair of heavy quark 
and antiquark at finite temperature, in terms of a quantum mechanical 
potential. We thus obtain the potential
by correcting {\em both the short and long-distance part} of the 
$Q \bar Q$ potential, with a dielectric
function, $ \epsilon(p)$~\cite{prc_vineet} embodying the effect of the 
medium
\begin{eqnarray}
\label{defn}
V(r,T)&=&\int \frac{d^3\mathbf p}{{(2\pi)}^{3/2}}
 \left( e^{i\mathbf{p} \cdot \mathbf{r}}-1 \right)~\frac{V(p)}{\epsilon(p)} ~.
\end{eqnarray}
We assume the same screening scales to regulate both terms (by multiplying 
with an exponential damping factor and is switched off after the Fourier 
transform is evaluated), to obtain the Fourier transform of the
potential\footnote{In Ref.~\cite{megiasind,megiasprd}, different scales 
for the Coulomb and linear pieces were employed through a dimension-two gluon
condensate.}:
\begin{equation}
\label{vk}
{V}(p)=-\sqrt{(2/\pi)} \frac{\alpha}{p^2}-\frac{4\sigma}{\sqrt{2 \pi} p^4}.
\end{equation}
We will now obtain the dielectric permittivity through the leading
anisotropic corrections to the self-energies and then to the 
static propagators in weak coupling HTL approximation.
In Keldysh representation, the retarded (R), advanced (A) and symmetric 
(F) propagators can be written as the linear combination of the components 
of the $(2 \times 2)$ matrix propagator in real-time formalism:
\begin{eqnarray}
\label{2a6}
   D_R^0 = D_{11}^0 - D_{12}^0 ~,~ D_A^0 = D_{11}^0 - D_{21}^0 ~,~
   D_F^0 = D_{11}^0 + D_{22}^0  ~,
\end{eqnarray}
where only the symmetric component involves the distribution functions 
and is of particular advantage for the HTL diagrams where the terms containing 
distribution functions dominate. Similar relations hold good for 
the retarded ($\Pi_R$), advanced ($\Pi_A$) and symmetric ($\Pi_F$) 
self energies.
Now the resummation of the propagators is done via the Dyson-Schwinger 
equation
\begin{eqnarray}
 {D}_{R,A}&=&D_{R,A}^0+D_{R,A}^0\Pi_{R,A}{D}_{R,A}~, \label{2b2}\\
 {D}_{F}&=&D_{F}^0+D_{R}^0\Pi _R{D}_{F}+D_F^0\Pi_{A} {D}_{A}+ 
 D_{R}^0\Pi _{F}{D}_{A}~. \label{2b7}
\end{eqnarray}
For the static potential, we need only the temporal component
(``00" $\equiv $ L) of the propagator, whose evaluation is
easier in the Coulomb gauge. Thus the above resummation (\ref{2b2})
can be recast through its temporal component as
\begin{eqnarray}
 D^L_{R,A(iso)}=D^{L(0)}_{R,A}+D^{L(0)}_{R,A}\Pi^L_{R,A(iso)}{D}^L_{R,A(iso)}~. \label{3b2}
\end{eqnarray}
The above relations are not satisfied for the anisotropic system 
due to the preferential direction of anisotropy. However, for 
anisotropic medium which exhibits a weak anisotropy ($\xi \ll 1$), we 
circumvent the problem, by expanding the 
propagators and self-energies in $\xi$:
\begin{equation}
 D=D_{\rm{iso}}+\xi D_{\rm{aniso}},\,\,\,\,\,\,\,\Pi=\Pi_{\rm{iso}}+\xi
\Pi_{\rm{aniso}} ~,\label{3b4}
\end{equation}
where the parameter $\xi$ is a measure of the anisotropy 
\begin{equation}
\xi = \frac{\langle \mathbf{p}_{T}^{2}\rangle}{2\langle p_{L}^{2}\rangle}-1~,~
\label{anparameter}
\end{equation}
where $ {p}_{L}= \mathbf{p}.\mathbf{n} $ and ${\bf p}_T =
\mathbf{p}-\mathbf{n}(\mathbf{p}.\mathbf{n}) $ are the components of momentum
parallel and perpendicular to the direction of anisotropy, $\mathbf{n}$,
respectively. 
Thus in the presence of small anisotropy, the (resummed) temporal component 
of the retarded (advanced) propagator becomes
\begin{equation}
D^L_{R,A(aniso)}= D^{L(0)}_{R,A}\, \Pi_{R,A (aniso)}^L{D}^{ L}_{R,A (iso)}+D_{R,A}^{L(0)}\,
 \Pi_{R,A (iso)}^L{D}^ L_{R,A (aniso)}\label{3b6}
\end{equation}

We will now calculate the temporal component of the retarded/advanced gluon 
self-energy in the HTL-approximation, where the leading isotropic 
contribution is 
\begin{eqnarray}
\Pi^{L}_{R,A(iso)}(P)=m_D^2\left(\frac{p_{0}}{2p}\ln\frac{p_{0}+p\pm i\epsilon}{p_{0}-p\pm i\epsilon}-1\right)~,
\label{iso}
\end{eqnarray}
with the prescriptions $+i\epsilon $ ($ -i\epsilon $), for the retarded 
and advanced self-energies, respectively and
$m_D^2$ (= $\frac{g^2 T^2}{6}(N_f+2 N_c$)) is the square of Debye mass. 
The full anisotropic contribution is then
\begin{eqnarray}
\Pi^{L}_{R,A(aniso)}(P)=\frac{m_D^2}{6}\left(1+3\cos 2\theta_p \right)
+\Pi_{R(iso)}^{L}(P)\left(\cos(2\theta_p)-\frac{{p_{0}}^{2}}{2p^{2}}
(1+3\cos 2\theta_p)\right)~,
\label{aniso}
\end{eqnarray}
Similarly the isotropic and anisotropic terms for the temporal component
of the symmetric self-energy are given by
\begin{eqnarray}
&&\Pi^{L}_{F(iso)}(P)=-2\pi i m_D^2\frac{T}{p}\Theta(p^2-{p_0}^2)~,\nonumber\\
&&\Pi^{L}_{F(aniso)}(P)=\frac{3}{2}\pi i m_D^2\frac{T}{p}
\left(\sin^2\theta_p 
+\frac{p_0^2}{{p}^2}~(3\cos^2\theta_p-1)\right)~\Theta(p^2-{p_0}^2).
\label{sym}
\end{eqnarray}
Thus the gluon self-energy is found to have both real and imaginary 
part which are responsible for the Debye screening and the Landau damping,
respectively where the former is usually obtained from the retarded and 
advanced self energy and the later is obtained from the symmetric self 
energy alone. 

Therefore the real part of the temporal component of 
retarded (or advanced) propagator in the static limit gives
\begin{eqnarray}
\Re D^{00}_{R,A}(0,p)=-\frac{1}{(p^2+m_D^2)}
+\xi \frac{m_D^2}{6(p^2+m_D^2)^2}\left(3\cos 2\theta_p-1 \right),
\label{rtrdprop}
\end{eqnarray}
and the static limit of the imaginary part of 
the temporal component of symmetric propagator is 
\begin{eqnarray}
\Im D^{00}_F (0,p)=\frac{-2\pi T m_D^2}{p(p^2+m_D^2)^2}
+\xi\left(\frac{3\pi T m_D^2}{2p(p^2+m_D^2)^2}\sin^2{\theta_p}
-\frac{4\pi T m_D^4}{p(p^2+m_D^2)^3} \left(\sin^2\theta_p-\frac{1}{3}\right)\right)
\label{f00}
\end{eqnarray}

We can now obtain the dielectric permittivity from the 
static limit of the ``00"-component of gluon propagator 
\begin{equation}
\epsilon^{{}^{-1}}(p)=-\lim_{\omega \to 0} {p^2} D_{11}^{00}(\omega, p)~,
\label{ephs}
\end{equation}
where the real and imaginary parts of $D^{00}_{11}$ can be written as
\begin{eqnarray}
\Re D^{00}_{11}(\omega,p)= \frac{1}{2}\left( D^{00}_{R}+D^{00}_{A}\right)
\label{R}~~ {\rm{and}}~~
\Im D^{00}_{11}(\omega,p)= \frac{1}{2} D^{00}_{F}.
\label{F}
\end{eqnarray}
The real-part of the potential is then obtained as
\begin{eqnarray}
\label{pot}
\Re V_{\rm(aniso)}({\bf r},\xi,T)&=&\int \frac{d^3\mathbf p}{{(2\pi)}^{3/2}}
 (e^{i\mathbf{p} \cdot \mathbf{r}}-1)
 \left(-\sqrt{(2/\pi)}\frac{\alpha}{p^2}- 
\frac{4\sigma}{\sqrt{2 \pi} p^4}\right) \times \nonumber\\
&&p^2\left[\frac{1}{(p^2+m_D^2)}-\frac{\xi m_D^2}{6(p^2+m_D^2)^2}
(3\cos(2\theta_p)-1)\right] \nonumber\\
&\equiv& \Re V_{1(aniso)}({\bf r},\xi,T)+ \Re V_{2(aniso)}({\bf r},\xi,T)~,
\end{eqnarray}
where $\theta_p$ is the angle between ${\bf r}$ and ${\bf n}$ (direction
of anisotropy) and $\Re V_{1(aniso)} ({\bf r},\xi,T)$ and 
$\Re V_{2(aniso)}(\mathbf{r}, \xi,T)$ are the 
medium modifications corresponding to the Coulomb and 
string term, respectively, are given by ($\hat r=rm_D$)
\begin{small}
\ba
\Re V_{1(\rm aniso)}(r,\theta_r,T)  &=&-\alpha m_D\left[\left( \frac{e^{-\hat{r}}}{\hat{r}}+1\right)+\xi \left[\left(\frac{e^{-\hat{r}}-1}{6}\right) \right.\right.\nonumber\\
&+&\left.\left.\left(\frac{e^{-\hat r}}{6}+\frac{e^{-\hat r}}{2\hat r}+\frac{e^{-\hat r}}{\hat{r}^2}+\frac{e^{-\hat r}-1}{\hat{{r}^3}}\right)(1-3\cos^2\theta_r)\right]\right]
\ea
\end{small}
and
\begin{small}
\ba
\Re V_{2(\rm aniso)}(r,\theta_r,T)  &=& \frac{2\sigma}{m_{{}_D}}\left[\left( \frac{e^{-\hat{r}}-1}{\hat{r}}+1\right)+2\xi\left[\left(\frac{e^{-\hat{r}}-1}{6\hat r}+\frac{e^{-\hat r}+2}{12}\right)   \right.\right.\nonumber\\
&+&
\left.\left.\left(\frac{e^{-\hat r}}{\hat r^2}+ \frac{5e^{-\hat r}+\hat re^{-\hat r}+1}{12\hat r}+\frac{e^{-\hat{r}}-1}{\hat{r}^3}   \right) (1-3\cos^2\theta_r)\right]\right]
\label{(eq:string)}
\ea
\end{small}
Thus the real-part of the potential in anisotropic medium becomes 
\begin{small}
\begin{eqnarray}
\label{fulrealpot}
\Re V_{\rm aniso}(r,\theta_r,T) &=&\frac{2\sigma}{m_{{}_D}}\left(\frac{e^{-\hat{r}}-1}{\hat{r}}+1\right) - \alpha m_D\left( \frac{e^{-\hat{r}}}{\hat{r}}+1\right)
+ \xi \frac{e^{-\hat{r}}}{\hat{r}}  \nonumber\\
&\times& 
\left[\frac{2 \sigma} {m_{D}}\left(\frac{e^{\hat{r}}-1}{\hat{r}^2}+\frac{\hat{r}^2e^{\hat{r}}-3}{3\hat r}-\frac{5 e^{\hat{r}}-\hat{r}+1}{12}\right) -\frac{\alpha m_{D}}{2}\left(\frac{e^{\hat{r}}-1}{\hat{r}^2}-\frac{1} {\hat{r}}-\frac{2\hat{r}e^{\hat{r}}-\hat r+3}{6} \right)\right.   \nonumber\\
&+&\left. \left[
\frac{2\sigma} {m_D}\left( 3 \frac{e^{\hat{r}}-1}{\hat{r}^2}-\frac{3}{\hat{r}}-\frac{e^{\hat{r}}+\hat r+5}{4} \right) -\frac{\alpha m_D}{2}\left(3 \frac{e^{\hat{r}}-1}{\hat{r}^2}-\frac{3}{\hat{r}}-\frac{\hat{r}+3}{2}\right) \right] \cos 2 \theta_r \right]\nonumber\\
&=& \Re V_{iso}(r,T) + V_{\rm{tensor}}(r,\theta_r,T). 
\label{fullp}
\end{eqnarray}
\end{small}
Thus the anisotropy in the momentum space introduces an angular ($\theta_r$)
dependence, in addition to the inter particle separation ($r$), to the 
potential, in contrast to the $r$-dependence only in an isotropic medium. 
The potential becomes stronger with the increase of anisotropy
because the (effective) Debye mass $m_D(\xi,T)$ in an anisotropic
medium is always smaller than in an isotropic medium. 
In particular, the potential for quark pairs aligned in the direction of 
anisotropy are stronger than the pairs aligned in the transverse direction.
\subsection{Imaginary part of the potential}
The imaginary part of the potential is obtained by the medium corrections 
to both the non-perturbative part (string term) and perturbative part
of the potential at T=0, by the imaginary part of the dielectric function (\ref{f00}):
\begin{eqnarray}
\Im V_{\rm(aniso)}({\bf r},\xi,T)&=&-\int \frac{d^3\mathbf{p}}{(2\pi)^{3/2}}
(e^{i\mathbf{p} \cdot \mathbf{r}}-1)
\left(-\sqrt{\frac{2}{\pi}}\frac{\alpha}{p^2}-\frac{4\sigma}{\sqrt{2\pi}p^4}\right)
p^2\left[\frac{-\pi T m_D^2}{p(p^2+m_D^2)^2}\right.\nonumber\\
&& \left.+\xi[\frac{3\pi T m_D^2}{4p(p^2+m_D^2)^2}\sin^2{\theta_p}
-\frac{2\pi T m_D^4}{p(p^2+m_D^2)^3}~(\sin^2\theta_p-\frac{1}{3})\right]
\nonumber\\
&& \equiv \Im V_{1(aniso)} ({\bf r},\xi,T)+ \Im V_{2(aniso)} ({\bf r},\xi,T) ~,
\end{eqnarray}
where $\Im V_{1(aniso)} ({\bf r},\xi,T)$ and 
$\Im V_{2(aniso)} (\mathbf{r},\xi,T)$ are the
imaginary contributions corresponding to 
the Coulombic and linear terms in anisotropic medium, respectively:
The contribution due to the perturbative term in the 
leading-order is given by~\cite{Dumitru:2007hy-09fy-09ni}
\begin{eqnarray}
\Im V_{1(aniso)}({\bf r},\xi,T)&= & 
-\alpha T \left( \phi_0(\hat{r})+
\xi\left[\phi_{1}(\hat{r}, \theta_r)
+\phi_2(\hat{r},\theta_r)\right] \right),                                     
\end{eqnarray}
where the functions $\phi_0(\hat{r})$, $\phi_{1}(\hat{r},\theta_r)$ and 
$\phi_{2}(\hat{r},\theta_r)$ are given by
\begin{eqnarray}
\phi_0(\hat{r})&=&-\alpha T\left(-\frac{{\hat{r}}^2}{9}
\left(-4+3\gamma_{E}+3\log\hat{r}\right)\right) \nonumber\\
\phi_{1}(\hat{r},\theta_r)&=& \frac{{\hat{r}}^2}{600} \left[123-
90 \gamma_{E}- 90\log\hat{r}     
+\cos 2\theta_r \left(-31+30\gamma_{E}+30\log\hat{r}\right)\right] \nonumber\\  
\phi_{2}(\hat{r},\theta_r)&=& \frac{{\hat{r}}^2}{90}(-4+3\cos 
2\theta_r)                
\end{eqnarray}
Similarly the imaginary part due to the nonperturbative (linear) term 
has also the isotropic and anisotropic term:
\begin{eqnarray}
\Im V_{2(aniso)}(r,\xi,T)= \frac{2\sigma T}{m_D^2} \left(  \frac{}{}
\psi_0(\hat{r})-\xi
\left[\psi_1(\hat{r},\theta_r)+\psi_2(\hat{r},\theta_r)\right] \frac{}{} 
\right)~,
\label{v2aniso}
\end{eqnarray}
where the functions $\psi_0(\hat{r})$, $\psi_1 (\hat{r},\theta_r)$ 
and $\psi_2 (\hat{r},\theta_r)$ are given by 
\begin{eqnarray}
\psi_0(\hat{r})&=&\frac{\hat r^2}{6}+\left(\frac{-107+60\gamma_E
+60\log(\hat r)}{3600}\right)\hat r^4+O(\hat r^5)~,\\
\psi_1(\hat{r},\theta_r)&=&\int \frac{dz}{z(z^2+1)^2}\left[1-\frac{3}{2}
\left(\sin^2\theta_r\frac {\sin{z\hat r}}{z\hat r}
+(1-3\cos^2\theta_r)G(\hat{r},z)\right)\right]~,\\
\psi_2(\hat{r},\theta_r)&=&-\frac{4}{3}\int\frac{dz}{z(z^2+1)^3}
\left[1-3\left[(\frac{2}{3}-\cos^2\theta_r)\frac {\sin{z\hat r}}{z\hat r}
+(1-3\cos^2\theta_r)G(\hat{r},z)\right]\right]
\end{eqnarray}
where 
\begin{eqnarray}
G(\hat{r},z)=\frac{z\hat r\cos(z\hat r)-\sin(z\hat r)}{(z\hat r)^3}
\end{eqnarray}
Thus the imaginary part of the potential in anisotropic medium in the 
leading logarithmic order becomes 
\begin{eqnarray}
\label{fullimgpot}
\Im V_{\rm{(aniso)}} (r,\theta_r,T)&=&-T\left(\frac{\alpha {\hat r^2}}{3}
+\frac{\sigma {\hat r}^4}{30m_D^2}\right)\log(\frac{1}{\hat r})\nonumber\\
&&+\xi T\left[\left(\frac{\alpha {\hat r^2}}{5}+\frac{3\sigma {\hat r^4}}{140m_D^2}\right)\right.
\left.-\cos^{2}\theta_r \left(\frac{\alpha {\hat r^2}}{10}+\frac{\sigma {\hat r^4}}{70m_D^2}\right)\right]\log(\frac{1}{\hat r})
\end{eqnarray}
where the magnitude is found to be smaller than the isotropic medium and 
decreases with the anisotropy.  In weak anisotropic limit, the imaginary 
part is a perturbation and thus provides an estimate for the (thermal) 
width for a particular resonance state: 
\begin{eqnarray}
\Gamma_{\rm(aniso)} &=& \int d^3 {\bf r}|\Psi(r)|^2\left[\alpha T{\hat r^2}
\log(\frac{1}{\hat r})\left(\frac{1}{3}-\xi
\frac{3-\cos 2\theta_r}{20}\right)\right.\nonumber\\
&&\left.+\frac{2\sigma T}{m_D^2}{\hat r^4}\log(\frac{1}{\hat r})
\frac{1}{20}\left(\frac{1}{3}-\xi\frac{2-\cos2\theta_r}{14}\right)\right]\nonumber\\
&=&T\left(\frac{4}{\alpha m_Q^2}+\frac{12\sigma}{\alpha^2m_Q^4}\right)\left(1-\frac{\xi}{2}\right)m_D^2 \log\frac{\alpha m_Q}{2m_D}~,
\end{eqnarray}
which shows that the in-medium thermal width in anisotropic medium becomes 
smaller than in isotropic medium and gets narrower with the increase of 
anisotropy. This is due to the fact
that the width is proportional to the square of the Debye mass and the
debye mass decreases with the anisotropy because the 
effective local parton density around a test (heavy) quark is smaller 
compared to isotropic medium.
\subsection{Dissociation in a complex potential}
In short-distance limit, the vacuum contribution dominates over the medium 
contribution even for the weakly anisotropic
medium and for the long-distance limit, the potential (\ref{fulrealpot}) in high temperature
approximation results a Coulomb plus a sub leading anisotropic contribution :
\begin{eqnarray}
\label{largp}
\Re V_{\rm{(aniso)}}(r,\theta_r,T) &\stackrel{\hat{r}\gg1}{\simeq}& -\frac{2\sigma}{m^2_{{}_D}r}
-\alpha m_{{}_D} -\frac{5\xi}{12}~\frac{2\sigma}{m^2_{{}_D}r} 
\left(1+\frac{3}{5}\cos 2\theta_r \right)\\
&\equiv & \Re V_{\rm{iso}} (\hat{r} \gg 1,T)+ V_{\rm{tensor}} (\hat{r} \gg 1~.
\theta_r,T)~,
\end{eqnarray}
where the anisotropic contribution 
($V_{\rm{tensor}} (\hat{r} \gg 1, \theta_r,T)$) is smaller than the isotropic 
one ($\Re V_{\rm{iso}} (\hat{r} \gg 1,T)$), so the anisotropic part
can be treated as perturbation. Therefore, the real part of binding energy 
may be obtained from the radial part of the Schr\"odinger equation
(of the isotropic component) plus the first-order perturbation 
due to the anisotropic component as :
\begin{eqnarray}
E_{\mathbf {bin}}^{\rm{aniso}} \stackrel{\hat{r}\gg1}{\simeq} 
\left( \frac{m_Q\sigma^2 }{m_{{}_D}^4 n^{2}} +
\alpha m_{{}_D} \right) +
\frac{2\xi}{3} \frac{m_Q\sigma^2 }{m_{{}_D}^4 n^{2}} ,
\end{eqnarray}
In the intermediate-distance scale, the real part of the potential 
(\ref{fulrealpot}) does not
look simple, the interaction becomes complex and needs to be
solved numerically. 
Usually the time- dependent or independent 
Schr\"odinger equation is solved by the finite difference time domain 
method (FDTD) or matrix method, respectively.
In the matrix method, the Schr\"odinger equation 
can be solved in a matrix form through a discrete basis, instead
of the continuous real-space position basis spanned by the states
$|\overrightarrow{x}\rangle$. Here the confining potential V is subdivided
into N discrete wells with potentials $V_{1},V_{2},...,V_{N+2}$ such that
for $i^{\rm{th}}$ boundary potential, $V=V_{i}$ for $x_{i-1} < x < x_{i};
~i=2, 3,...,(N+1)$. Therefore for the existence of a bound state, there
must be exponentially decaying wave function
in the region $x> x_{N+1}$ as $x \rightarrow  \infty $ and
has the form:
\begin{equation}
\Psi_{N+2}(x)=P_{{}_E} \exp[-\gamma_{{}_{N+2}}(x-x_{N+1})]+
Q_{{}_E} \exp [\gamma_{{}_{N+2}}(x-x_{N+1})] ,
\end{equation}
where, $P_{{}_E}= \frac{1}{2}(A_{N+2}- B_{N+2})$,
$Q_{{}_E}= \frac{1}{2}(A_{N+2}+ B_{N+2}) $ and,
$ \gamma_{{}_{N+2}} = \sqrt{2 \mu(V_{N+2}-E)}$. The eigenvalues
can be obtained by identifying the zeros of $Q_{E} $.
We have then obtained the real and imaginary part of 
the binding energies of the bottomonium states (shown in Fig 1),
which is found to increase with the anisotropy.

We now study the dissociation of the resonances 
when the binding energy decreases with the increase of the 
temperature and becomes equal to $\sim \Gamma$~\cite{Mocsy05-08,Burnier07}.
The dissociation temperatures ($T_D$'s) can also be obtained 
from the intersection of the binding 
energies obtained from the real and imaginary part of the 
potential~\cite{Strickland:2011aa,Margotta:2011ta}, respectively. 
The $T_D$'s for the $\Upsilon$ (1S) and $\Upsilon$ (2S) states are 
$1.97~T_c$ and $1.44T_c$, respectively (Table 1) in isotropic medium 
($\xi=0$) and increases with the increase of anisotropy ($\xi>0$),
{\em i.e.} the bottomonium states persist higher
temperatures (2.1 $T_c$ for $\xi=0.6$) in a anisotropic plasma, 
which can be parametrized as
$T_{\rm{aniso}}^D(\xi) \simeq T_{\rm{iso}}^D\left(1+\frac{\xi}{7}\right)$,
compared to the relation $T_{\rm{aniso}}^D(\xi) =
T_{\rm{iso}}^D\left(1+\frac{\xi}{6}\right)$ by Laine et al.\cite{Burnier:2009yu}.
Our results are found relatively higher
compared to similar calculation~\cite{Strickland:2011aa,Margotta:2011ta},
which may be due to the absence of three-dimensional medium modification
of the linear term in their calculation.
\begin{figure}[h]
\vspace{1.75in} 
\begin{center}$
\begin{array}{cc}
\includegraphics[width=2.71in,height=2.5in]{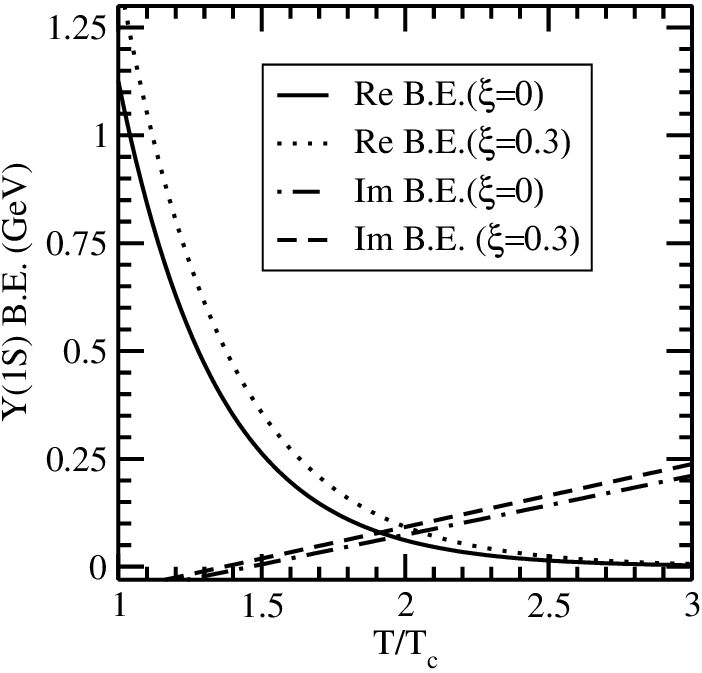} & \hspace{0.25in}
\includegraphics[width=2.5in,height=2.5in]{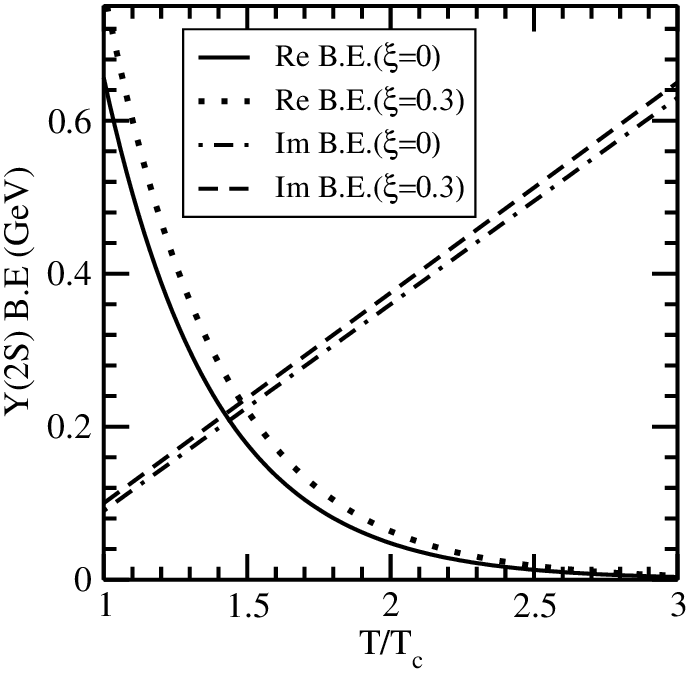} \\ 
\end{array}$
\end{center}
\caption{\footnotesize The real and imaginary part of the binding energies
for the 1S and 2S states.}
\end{figure}
\begin{table}[h] 
\centering
\begin{tabular}{|c|c|c|c|c|c|c|c|c|}
\hline
\multirow{2}{*}{State}& \multicolumn{3}{c|}{$T_D$} & \multirow{2}{*} \large{$\tau_F$}& \multirow{2}{*}\large{$\eps$}&\multirow{2}{*}\large {$c_{s}^2$}\\
\cline{2-4}
&$\xi=0$ &$\xi=0.3$&$\xi=0.6$& \small{(fm)}& \small{($\xi=0$)} &\\
\hline\hline
$\Upsilon$(1S) & 1.97&2.01&2.11& 0.2&42.49&0.307 \\
\hline
$\Upsilon$(2S)  &  1.44&1.50&1.54& 0.4&12.74&0.284\\
\hline
$\Upsilon$(3S) & 1.12&1.15&1.21& 0.6&5.21&0.249\\
\hline
$\chi_{b1}$& 1.57&1.59&1.64&0.4 &17.66& 0.292\\
\hline
\end{tabular}
\caption{\footnotesize Dissociation temperatures ($T_D$) in units of $T_c$
for bottomonium states at different anisotropies ($\xi$)
along with their (three sets) formation times($\tau_F$),
screening energy densities ($\eps$) and square of speed of sound($c_s^2$) in isotropic medium. }
\end{table}
\section{Quarkonium in expanding medium}
Let us now consider a nucleus-nucleus collision, where the partons
 are formed
at time $\tau_i \sim Q_s^{-1}$ ($Q_s$ is the saturation
scale) and the system started evolving. There may be three plausible
scenarios in space-time evolution {\em viz.} i) the
partons are initially isotropized ($\tau_i=\tau_{iso}$), i.e.,
the system evolves hydrodynamically, ii) the system never isotropizes
($\tau_{iso} \rightarrow \infty$), i.e., it undergoes free streaming
motion and iii) finally the system takes finite time to isotropize 
($\tau_{i}<\tau<\tau_{iso}$), i.e., undergoes 
through successive anisotropic (pre-equilibrium) and 
isotropic (equilibrium) phases. The pre-equilibrium
era may be conceived by the fact 
that the asymptotic weak-coupling at the early stage of the collision 
enhances the expansion in the beam direction (longitudinal) 
substantially than the 
radial expansion and results an anisotropy.
This anisotropy makes the partonic system unstable with respect to 
the chromo magnetic plasma modes~\cite{Romatschke}, which facilitate 
the system to isotropize quickly~\cite{Arnold05,Mrowezynski93,Arnold03}.
Recently there have been significant advances in the dynamical models 
for the plasma evolution to incorporate the momentum anisotropy 
~\cite{Mrowczynski:2000ed,Strickland:2007fm,martinez-strickland}.

Let us consider a region of energy density in the transverse plane, 
which is 
greater than or equal to the screening energy density, $\epsilon_s$ 
($\propto T_D^4$).
During the expansion, if the system has been cooled to an energy density
less than or equal to $\epsilon_s$, the $Q \bar Q$ pair would escape
and form the (quarkonium) resonance. On the other hand if the energy
density is still higher than $\epsilon_s$, the resonance
will not form and suppress the quarkonium production.
Thus the pattern of suppression of the quarkonium states 
depends on how rapidly the system cools and how large is the
the screening energy density of the particular resonance state. 
The former depends on how to model the evolution of the system,
where the equation of state needed to close the hydrodynamic equations 
is still not clear and how to incorporate the dissipative 
forces in the dynamics. The later (the screening energy density) depends on the
properties of quarkonium states in the medium (which may or may not be
isotropic). However, the (finite) formation time and the intrinsic 
transverse momentum of the resonance enrich the suppression pattern 
more interesting.

Thus the discussion on the suppression of resonances in expanding medium 
is three fold respects: First we discuss on the equation
of state (EOS) which gives the speed of sound
(which controls the expansion of the medium) as a
function of temperature, in contrast to the constant value usually adopted
in the literature and use the EOS to evaluate the screening energy density 
corresponding to the temperature $T_D$.
Secondly we discuss the evolution of the system first through
the pre-equilibrium era and subsequently the (local) equilibrium era
in the Bjorken boost-invariant expansion in the 
presence of dissipative forces in the stress tensor.
Finally the above ingredients are coupled to quantify the 
suppression of the bottomonium states at the LHC.

\subsection{Lattice equation of state}
The pressure is the primary observable to study 
the QCD equation of state which, at finite chemical potential ($\mu_i$), can be written 
through the Taylor-expansion~\cite{Borsayni}:
\ba
\frac{p(T,\{\mu_i\})}{T^4} = \frac{p(T,\{0\})}{T^4} + \frac{1}{2} \sum_{i,j} \frac{\mu_i\mu_j}{T^2} \chi_2^{ij},
\label{eq:pmu}
\ea
with the susceptibilities
\be
\chi_2^{ij} \equiv \frac{T}{V} \frac{1}{T^2}\left.\frac{\partial ^2 \log\cal{Z}}{\partial \mu_i \partial \mu_j}\right|_{\mu_i=\mu_j=0}.
\label{eq:chi2def}
\ee
The trace anomaly, $I(T,\mu)$ is another quantity of interest in equation of
state which can be obtained from the relation: 
\ba
\frac{I(T,\mu)}{T^4}\equiv \frac{ \epsilon(T,\mu)-3p(T,\mu)}{T^4}
=\frac{I(T,0)}{T^4} \,+\, \frac{\mu^2}{2T} 
\frac{\partial \chi_2}{\partial T}
\label{eq:imu}
\ea
In the limit of vanishing baryon chemical potential, the trace anomaly 
and the Taylor-coefficients (susceptibilities) was parametrized~\cite{Borsayni} as 
\begin{eqnarray}
\frac{I(T)}{T^4}& = &e^{-h_1/t-h_2/t^2}\left[h_0+ \frac{f_0~[\tanh(f_1 ~t+f_2)+1]}{1+g_1~t+g_2~t^2}\right]~.\\
\chi_2(T)&=&e^{-h_3/t-h_4/t^2} f_3~[\tanh (f_4~t+f_5)+1]~,
\end{eqnarray}
where $t=T/200$ and the values of other parameters are given 
in~\cite{Borsayni}.

The inverse relation between the pressure and the trace anomaly at $\mu=0$ 
then becomes
\be
\frac{p(T,0)}{T^4} = \int_0^Td T' \frac{I(T',0)}{T'^5}.
\label{eq:invpi}
\ee
The energy density $\epsilon$ is then obtained from the trace 
anomaly and the pressure
\be
\epsilon=I+3p~,
\ee
hence the speed of sound $c_s$ can  be obtained from the relation:
\be
c_s^2= \left.\frac{\partial p}{\partial \epsilon}\right|_{s/n}~.
\label{eq:cs2}
\ee
  We have shown that how the speed of sound varies with temperature rapidly
in the vicinity of the critical point
and approaches towards the asymptotically ideal value (1/3) (in Fig.2) for
very high temperature. In
particular we have marked the values of $c_s^2$'s at the dissociation
temperatures ($T_D$'s) of the $\Upsilon$ (nS) states and indicates that
how the expansion of the system deviates from the ideal one 
at $T_D$'s and hence has a large bearing on the suppression.

Thus the equation of state can be used to calculate the energy 
density ($\epsilon_s$) at the dissociation temperature ($T_D$) and also
be used as an input to the hydrodynamics equation of motion.
Another important quantity in the quarkonium suppression is the screening 
time, which can also be obtained from the screening energy density 
$\epsilon_s$ and the speed of sound.
\begin{figure}[]
\vspace{0.1in} 
\begin{center}
\includegraphics[width=2.5in,height=2.5in]{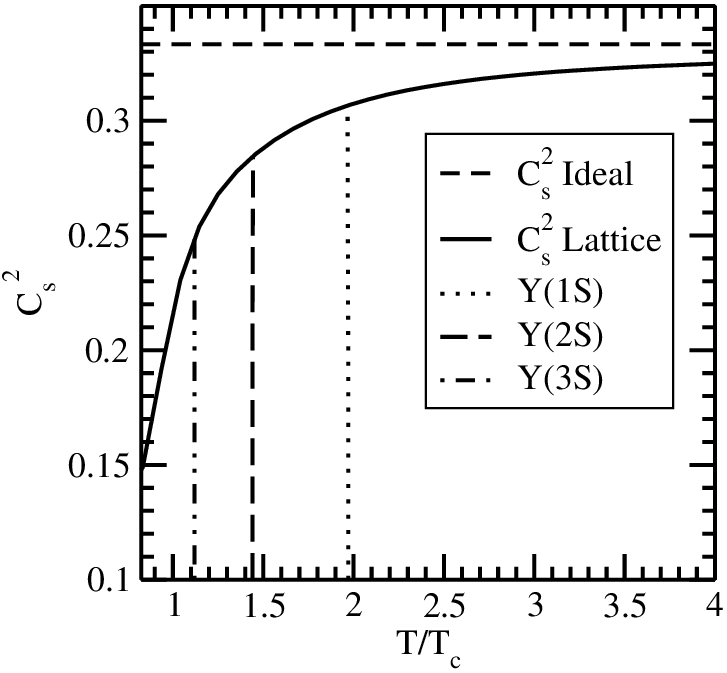}
\end{center}
\caption{\footnotesize variation of the speed of sound}
\end{figure}
\subsection{Evolution in pre-equilibrium era}
The evolution of the pre-equilibrium (anisotropic) hydrodynamics may be 
dealt in two ways: the first one is a phenomenological and 
refers directly to tensor structure of an anisotropic
fluid~\cite{Ryblewski:2010bs,Ryblewski:2012rr,Florkowski:2010cf} and the 
second one employs the transport equation for the gluon distribution
function in the anisotropic background
\cite{Martinez:2010sd-12tu}. Phenomenologically the  
generalized anisotropy parameter can be written in (1+1) dimension
 \be 
 \xi(\tau,\delta)=\left(\frac{\tau}{\tau_{i}}\right)^\delta-1 ~,
 \label{eq:genzeta}
 \ee
where the interpolating co-efficient, $\delta$ characterizes the 
various isotropization process, {\em viz.} the asymptotic limits 
$\delta \rightarrow 0$ and $2$ represent the (local equilibrium)
hydrodynamics and the free-streaming, respectively whereas
the intermediate values $1/6 \le \delta  \le 1/2 $ and $2/3$ denote 
the plasma instability and the collisional broadening, respectively.
For the general value of  $\delta$, the proper time dependence  of the 
energy density, the hard momentum scale  and the number density 
for large times, $\tau \gg \tau_{i}$,  can be written as,
\ba 
\varepsilon(\tau) &=&  \varepsilon_{0} \left( \frac{\tau_{i}}{\tau}\right)^{4(1-\delta/8)/3},\\
p_{\rm hard}(\tau) &=& T_{0}\left( \frac{\tau_{i}}{\tau}\right)^{(1-\delta/2)/3},\\
n(\tau)&=&n_{0}\left( \frac{\tau_{i}}{\tau}\right)~,
\ea 
respectively. The smoothness of the transition from a non zero 
value (of $\delta$) to 0 at $\tau \sim \tau_{iso}$, is governed by
a smeared  step function $\lambda(\tau)$ \cite{Martinez:2010sd-12tu,martinez-strickland}, 
\begin{equation}
 \lambda(\tau) = \frac{1}{2} \left[{{ 
\tanh}{\left(\frac{\gamma (\tau-\tau_{\rm iso})}{\tau_i }\right)+1}}\right]~,
\label{eq:lambda}
\end{equation}
where the parameter,  $\gamma$ sets the sharpness of the transition 
from pre-equilibrium to (local equilibrium) hydrodynamic behavior. Thus
the above dependence, in terms of $\lambda (\tau)$, become
\ba 
\xi(\tau,\delta) &=&\left(\frac{\tau}{\tau_i}\right)^{\delta(1-\lambda(\tau))}-1
 \\
{\cal E}(\tau)&=&{\cal E}_{\rm 0} {\cal R({\xi})}\
{\cal \bar U}^{4/3}\label{eq:edpreq}\\
p_{\rm hard}(\tau) &=&T_{0} ~\ {\cal\bar U}^{1/3}~ 
\label{eq:modelEQs} 
\ea
where the functions ${\cal R({\xi})}$ and ${\cal \bar U}$ are given by
\ba
{\cal R({\xi})}&=&\frac{1}{2} \left( \frac{1}{\xi+1}+\frac{\tan^{-1}\sqrt{\xi}}{\sqrt{\xi}}\right),\\ 
{\cal \bar U}& = &{\cal U}(\tau)/  {\cal U}(\tau_ {i}) ,
\ea
with 
\ba
{\cal U}(\tau)  &\equiv &  \left[{\cal R} \left((\frac{\tau_{iso}}{\tau_{i}})^{\delta}-
1\right)\right]^{3\lambda(\tau)/4}  \left(\tau_{\rm iso}/\tau\right)^{1-\delta(1-\lambda(\tau))/2}. 
\label{eq:ubar}
\ea
\begin{figure}[]
\vspace{0.1in} 
\begin{center}$
\begin{array}{cc}
\includegraphics[width=2.5in,height=2.5in]{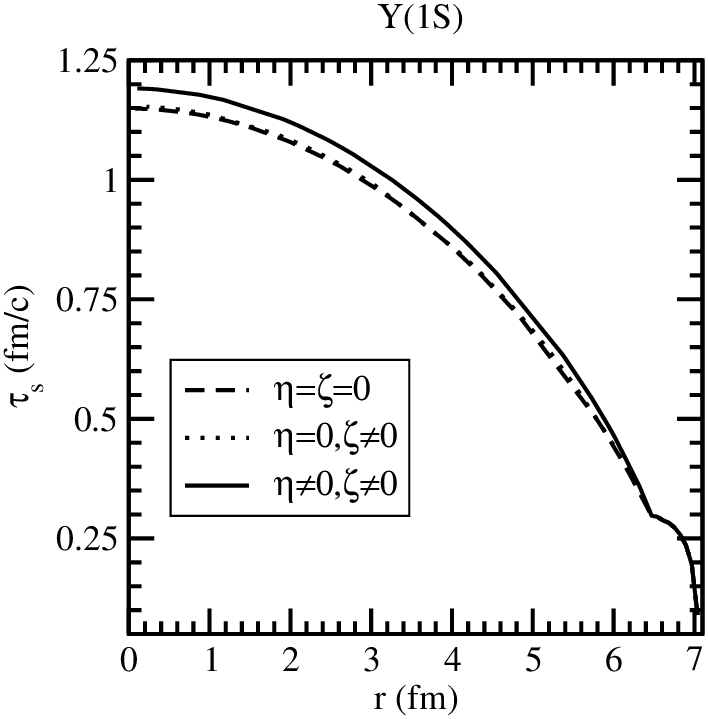} & \hspace{0.25in}
\includegraphics[width=2.5in,height=2.5in]{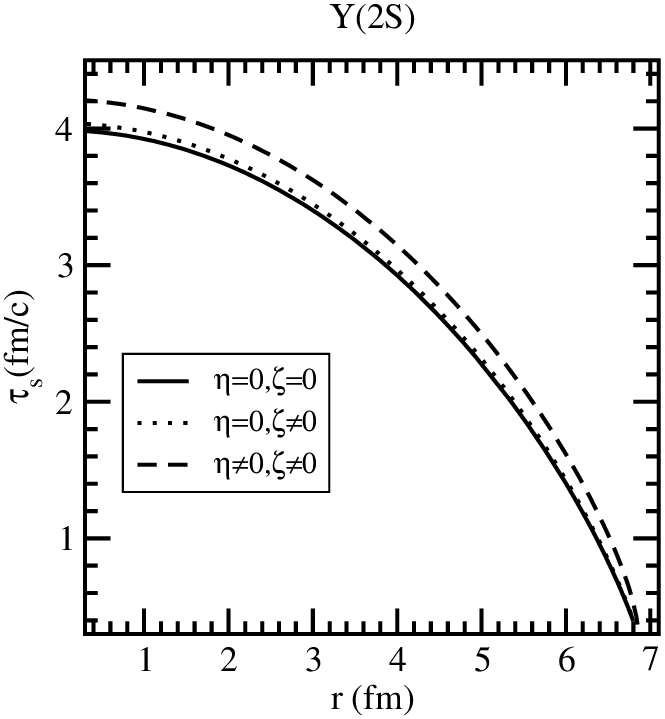} \\ 
\end{array}$
\end{center}
\caption{\footnotesize Constant energy density contour for 
the $\Upsilon$ (1S) (left panel) and  the $\Upsilon$(2S) (right panel)
for different values of shear ($\eta$) and bulk ($\zeta$) viscosities.}
\end{figure}
\begin{figure}[h]
\vspace{.1in}
\begin{center}$
\begin{array}{cc}
\includegraphics[width=2.5in,height=2.5in]{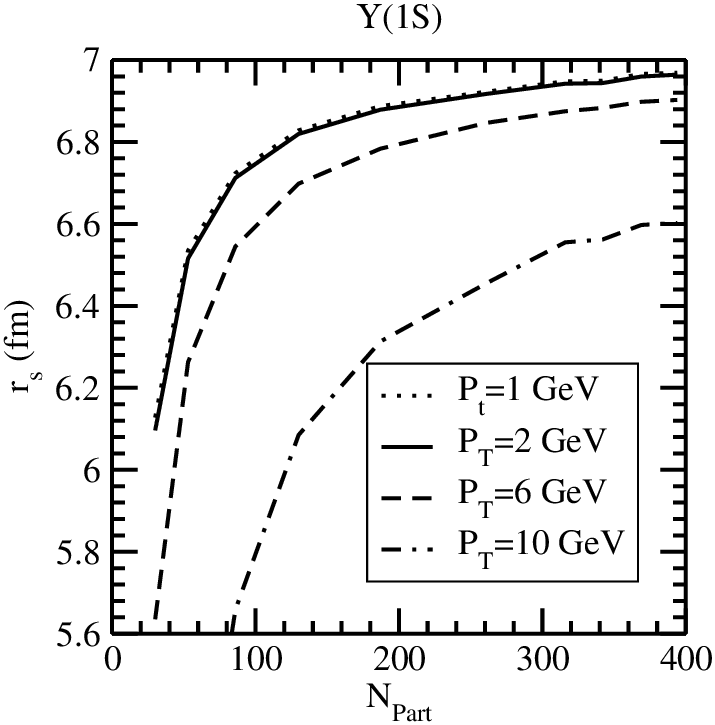} & \hspace{0.25in}
\includegraphics[width=2.5in,height=2.5in]{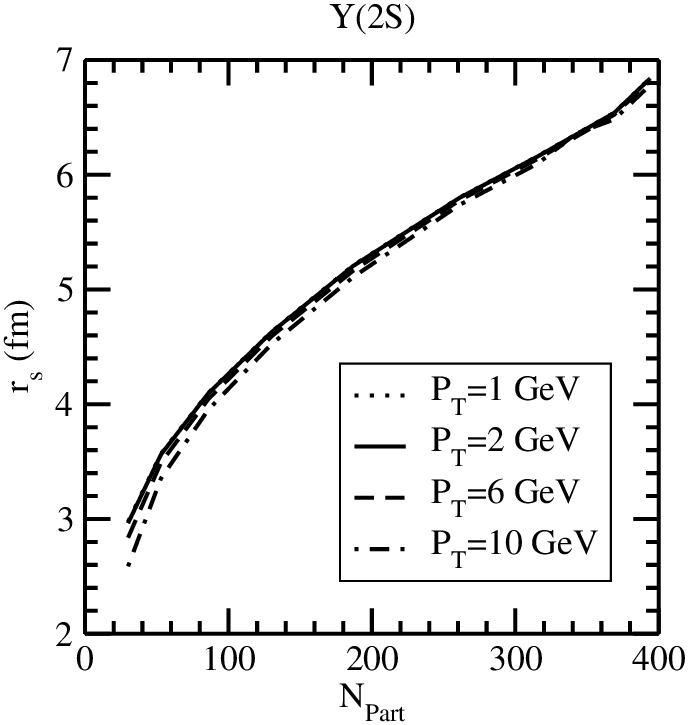} \\ 
\end{array}$
\end{center}
\caption{\footnotesize Centrality dependence of  screening radius at fixed $P_T$ 
for $\Upsilon$(1S)(Left panel) and $\Upsilon$(2S) (right panel).}
\end{figure}
\subsection{Evolution in equilibrium era: Bjorken expansion }
When the rate of interaction overcomes the expansion rate
the system attains local thermodynamic equilibrium for 
times $\tau \geq \tau_{iso}$. 
The energy momentum tensor of the plasma in the absence of dissipative 
forces is written as:
\begin{equation}
T^{\mu\nu}= (\epsilon+p)u^\mu u^\nu + g^{\mu \nu} p~,
\label{tmu}
\end{equation}
where $\epsilon$ and $p$ are the energy density and the pressure, respectively.
Then the Bjorken's boost-invariant longitudinal expansion
gives the equation of motion:
\begin{equation}
\frac{d \epsilon}{d \tau} =-\frac{\epsilon+p}{\tau}~,
\label{bj}
\end{equation}
where the equation of state ($p=c_s^2 \epsilon$) has been coupled 
to give rise 
\begin{eqnarray}
\label{eq0}
\epsilon(\tau) \tau^{1+c_s^2}= \epsilon(\tau_i)\tau_i^{1+c_s^2}.
\end{eqnarray}
Now we study the corrections to the Bjorken expansion
due to the dissipative forces in the energy-momentum tensor:
\begin{equation}
T^{\mu\nu}= (\epsilon+p)u^\mu u^\nu + g^{\mu \nu} p +\Pi^{\mu\nu} ~,
\label{tmun1}
\end{equation}
where the dissipative part (viscous stress tensor), $\Pi^{\mu\nu}$ is given by 
\be
\Pi^{\mu\nu}=\eta \left( \nabla^\mu u^\nu + \nabla^\nu u^\mu -
\frac{2}{3} \nabla^{\mu \nu} \nabla^\rho u_\rho \right)
+ \zeta \nabla^{\mu \nu} \nabla^\rho u_\rho ~,
\ee
where $\eta$ and $\zeta$ are the shear and bulk viscosities, respectively
and $\nabla^\mu=\nabla^{\mu \nu} \partial_\nu$ with $\nabla^{\mu \nu}
=g^{\mu \nu} - u^\mu u^\nu$.
In first-order viscous hydrodynamics, the bulk and shear stresses can be 
written in a gradient expansion:
\be
\Pi=-\zeta \partial^\mu u_\mu, ~~~ \pi^{\mu\nu}=\eta \langle 
\nabla^\mu u^\nu \rangle, 
\ee
where $ \langle \nabla^\mu u^\nu \rangle$ is the symmetrized velocity 
gradient. The Israel-Stewart theory of second-order
dissipative hydrodynamics \cite{Israel:1979wp} modifies
the equation of motion for the ideal fluid (\ref{tmun1}) 
into~\cite{Heinz:2005zi,Muronga:2003ta,Baier:2006um,Baier:2007ix} 
\be
\frac{d \epsilon} {d \tau}=-\frac{1}{\tau}(\epsilon+p-\Phi +\Pi)
~,
\label{eq5}
\ee
where the bulk ($\Pi$) and the shear stress ($\Phi$) will asymptotically
(after the relaxation times $\tau_\Pi$ and $\tau_\pi$, respectively)
reduce to their first-order values. In the Navier-Stokes limit,  the 
one-dimensional boost-invariant expansion gives~\cite{Fries:PRC782003}, 
\begin{equation}
\Phi = \frac{4\eta}{3\tau} ,\qquad \Pi = - \frac{\zeta}{\tau} \, .
\label{eq:eta-zeta}
\end{equation}
Substituting  the values of $\Phi$ and $\Pi$ in ~(\ref{eq5}), the
Bjorken longitudinal expansion can be read as:
\be
\frac{d \epsilon} {d \tau}+\frac{\epsilon+p}{\tau}=
\frac{\frac{4\eta}{3}+\zeta}{\tau^2}~,
\label{eqbj2}
\ee
whose solution can be obtained with the EoS $p=c_s^2 \epsilon$,
\begin{eqnarray}
\label{eqs1}
\epsilon(\tau) \tau^{1+c_s^2}+c\left[\frac{4\eta}{3s}+\frac{\zeta}{s}\right] \frac{\tau^{1+c_s^2}}{{\tilde{\tau}}^2}
&=&\epsilon(\tau_i)\tau_i^{1+c_s^2}
+c\left[\frac{4\eta}{3s}+\frac{\zeta}{s}\right] \frac{\tau_i^{1+c_s^2}}{ {\tilde{\tau_i}}^2} 
\end{eqnarray}
where the constant, $c$ is $(1+c_s^2)a_f T_{i}^3\tau_{i}$ with 
$a_f=(16+21n_f/2)\pi^2 /90$, and ${\tilde{\tau}}^2$ 
(or ${\tilde{\tau}}_i^2$) are denoted by $(1-c_s^2)\tau^2$ 
($(1-c_s^2)\tau_i^2$), respectively.
The first term in both LHS and RHS accounts for the contributions coming 
from the zeroth-order expansion while  the second term is due to the  
viscous corrections. 

In the present work we use the shear viscosity to-entropy
ratio, $\eta/s$ from the perturbative QCD~\cite{pqcd} and
AdS/CFT calculations~\cite{ADS-kovtun-dtson-PRL942005},
whereas for the bulk-viscosity, $\zeta/s$ we consider the 
parametrization 
in~\cite{jaiswal:PRC872013,rajgopal-jhep2010,Mayer-PRL2008}, which 
suggest a sharp peak in the vicinity of $T_c$ and is tiny 
below $T_c$~\cite{mprakash-physrepo-1993}:
\begin{equation}
\label{eq:zetas}
\zeta{\mbox{/s}}=\left\{ \begin{array}{ll}
a_1\exp \left(\frac{T-T_c}{\Delta T}\right)+b_1\left(\frac{T_c}{T}\right)^2 & \mbox{if $T > T_c$} \\
a_1\exp \left(\frac{10 (T_c-T)}{\Delta T}\right)+\frac{b_1}{10}\left(\frac{T}{T_c }\right)^2 & \mbox{if $T_c \ge T$}~,
\end{array}
 \right.
\end{equation}
where the parameter $a_1$ (=0.901) and the $\Delta$T (=$T_c/14.5$) 
controls the height and the width of the curve, both of which are not 
well understood and may be varied considerably. The parameter $b_1$ (=0.061)
is obtained by fitting Meyer’s central value of~$\zeta/s$ at higher 
temperatures~\cite{jaiswal:PRC872013}.
\subsection{Survival of $b \bar b$ states}
Let us take a simple parametrization for the initial energy density profile on
the transverse plane:
\begin{equation}
\label{eq:edprofile}
\epsilon(\tau_i,r)=\epsilon_i A_{{}_T}(r)~,
\end{equation}
with the profile function
\begin{equation}
A_{{}_T}(r) =\left(1-
\frac{r^2}{R_T^2}\right)^{\beta} \theta(R_T-r)
\end{equation}
where $r$ is the transverse co-ordinate, $R_T$ is the transverse 
radius of the nucleus, and $\beta$ represents the proportionality of 
the deposited energy to the nuclear thickness. 
Thus the average initial energy density, $\langle 
\epsilon_i \rangle$ 
 can be obtained as
\begin{equation}
\epsilon_i=(1+\beta)  \langle \epsilon_i \rangle~,
\end{equation} 

With this initial energy density profile (\ref{eq:edprofile}), we now obtain 
the  screening time ($\tau_s (r)$), when the energy density 
drops to the screening energy density $\epsilon_s$, and construct the 
screening energy density contour. Since the system evolves through the 
successive pre-equilibrium and equilibrium era, so the entire contour 
is obtained by amalgamating the contours in pre-equilibrium and 
the equilibrium era. The contour in the 
pre-equilibrium era is obtained from the energy density (\ref{eq:edpreq}) 
whereas the equilibrium era gives the contour:
\begin{eqnarray}
\label{taus}
\tau_s(r)=\tau_i \left(\frac{\tilde \tau_s}{\tilde \tau_i}\right)^{2/1+c_s^2}{\bigg[ \frac{\epsilon_i(r) \tilde\tau_{i}^2 +c(\frac{4\eta}{3s}+\frac{\zeta}{s})}
{\epsilon_s \tilde{\tau}_s^2 +c(\frac{4\eta}{3s}+\frac{\zeta}{s})}
\bigg ]}^{1/1+c_s^2}~,
\end{eqnarray}
The significance of the contour can be understood as follows: If a 
$Q \bar Q$ pair 
is produced inside the contour, the pair cannot escape and hence the 
resonance cannot be formed. If it is produced outside the contour, it 
survives. Since the $Q \bar Q$ pair takes finite time ($\tau_F$) to 
form the physical resonances ($J/\psi$, $\Upsilon$ etc.),
the boundary of the region ($r_s$), where the quarkonium formation is 
suppressed, has been quantified by equating the duration of screening
time $\tau_s(r)$ to the formation time $t_F$ in the plasma 
frame (=$\gamma \tau_F$, where $\gamma$ is the dilation factor).
For the equilibrium era undergoing through 
the Bjorken boost-invariant expansion, the 
screening radius can calculated as:
\begin{eqnarray}
\label{rs}
r_s&=& R_T { \left( 1- A \right)}^{\frac{1}{2}}~\Theta \left( 1-A \right)~,\\
A &=& \bigg[ \frac{\epsilon_s}{\epsilon_i} 
\left({\frac{t_F}{\tau_i}} \right)^{1+c_s^2} 
+ \frac{c\left({\frac{4\eta}{3s} + \frac{\zeta}{s}}\right)}{\epsilon_i} \bigg(
 \frac{(\frac{ t_F}{\tau_i})^{1+cs^2}}{{\tilde{\tau}}_s^2}-\frac{1}{{\tilde{\tau}}_i^2}
 \bigg)  \bigg]^{1/\beta},
\end{eqnarray}
which depends on the initial conditions, the dynamics of the evolution, and
also the dynamical properties of the resonance states. Since the 
initial conditions are related to the centrality of the collisions, thus 
the screening boundary gives rise a centrality dependent suppression pattern.
Suppose a $Q \bar Q$ pair is created initially at $\mathbf{r}_0$ 
with the transverse momentum $\mathbf{p}_T$ on the
transverse plane. By the time the resonance is formed, the pair moves then 
to a new position ${\bf r}={\bf r}_0+ t_F \mathbf{p}_T/M$
and if $|{\bf r}|$ is greater than or equal to the screening radius $r_{s}$,
the pair will escape the deadly contour, otherwise
the resonance will never be formed.
This gives rise a characteristic dependence of  $p_{{}_T}$ in the
suppression pattern as well as the inequalities of 
the cosine of the angle between $\mathbf{r}$ and $\mathbf{p}_T$ vectors:
\begin{equation}
\cos \phi\,\geq\,\left[(r_s^2-r^2)\,M-\tau_F^2\,p_T^2/M\right]/
\left[2\,r\,\tau_F\,p_T\right],
\label{cosphi}
\end{equation}
which leads to a range of values of $\phi$ when the quarkonium would
escape. 

Now we can write for the survival probability of the quarkonium:
\begin{eqnarray}
\spt=\left[\int_0^{R_T} \, r \, dr \int_{-\phi_{\mbox{max}}}
^{+\phi_{\mbox{max}}}\,
d\phi\, P(\mathbf{r},\mathbf{p}_T)\right]/
\left[2\pi \int_0^{R_T} \, r\, dr\, P(\mathbf{r},\mathbf{p}_T)\right],
\label{sspt}
\end{eqnarray}
where $\phi_{\mbox{max}}$ is the maximum positive angle
($0\leq \phi \leq \pi$)
allowed by Eq.(\ref{cosphi}):
\begin{equation}
\phi_{\mbox{max}}=\left\{ \begin{array}{ll}
\pi & \mbox{if $y\leq -1$}\\
\cos^{-1} |y| & \mbox{if $-1 < y < 1$}\\
 0          & \mbox{if $y \geq 1$}
 \end{array}
 \right .,
\end{equation}
with 
\begin{equation}
y= \left[(r_s^2-r^2)\,M-\tau_F^2\,p_T^2/M\right]/
\left[2\,r\,\tau_F\,p_T\right],
\end{equation}
$M$ is the mass of the resonance and $P$ is the probability for the 
quark-pair production at
($\mathbf{r}$, $\mathbf{p}_T$), in a hard collision which
may be factored out as
\begin{equation}
P(\mathbf{r},\mathbf{p}_T)=f(r)g(p_T),
\end{equation}
where we take the profile function f(r) as
\begin{equation}
f(r)\propto \left[ 1-\frac{r^2}{R_T^2}\right]^\alpha \theta(R_T-r)~.
\end{equation}

Often experimental measurement of survival probability at a given
number of participants ($N_{{}_{\rm part}}$) or rapidity ($y$)
is reported in terms of the $\pt$-integrated yield ratio:
\begin{equation}
\sinpt  = \frac{\int_{\ptmin}^{\ptmax}  d \pt S(\pt)}
{\int_{\ptmin}^{\ptmax} d \pt}.
\end{equation}
The production of $b \bar b$ mesons occur in-part
through the production of higher excited $b\bar b$ states
and their decay into the ground state. Since the ground
and excited states have different sizes (binding energies), 
the excited states will dissolve  earlier compared to the tightly bound 
ground states, so a sequential suppression results. However, the situation 
may not be that simple
because the states have different formation times too, opposite to
their binding energies.
So while calculating the $\pt$-integrated inclusive
survival probability for individual  states, the feed-down corrections 
may be taken into account as:
\begin{eqnarray}
\sincl (3S) &=& {\sinpt}(3S),\\
\sincl (2S)&=& f_1 {\sinpt} (2S) +f_2 \sinpt (3S),\\
\sincl (1S)&=& g_1 {\sinpt} (1S) +g_2 \sinpt 
\chi_{b1} +g_3 \sinpt (2S)+ g_4 \sinpt (3S),
\end{eqnarray}
where the branching factors $f_i$'s and $g_i$'s are obtained 
from the CDF measurement~\cite{CDF:Collab}, where $g_{i}$'s are 0.509, 0.271, 
0.107 and 0.113, respectively, assuming the survival probabilities of 
$\Upsilon(3S)$ and $\chi_b$(2P) are same with $g_4$ as combined fraction 
while factors $f_1$ and $f_2$ are taken as 0.5.
\begin{figure}[]
\vspace{.11in}
\begin{center}$
\begin{array}{cc}
\includegraphics[width=2.5in,height=2.5in]{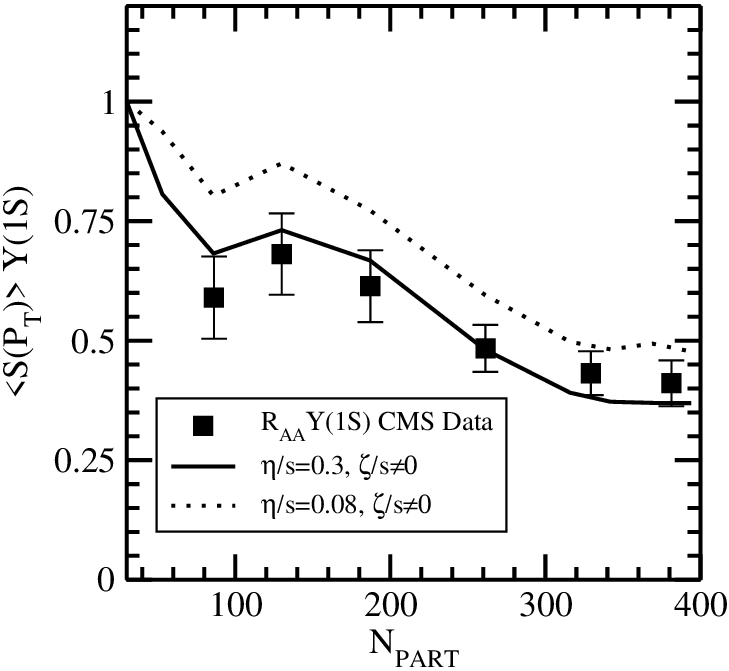} & \hspace{0.25in}
\includegraphics[width=2.5in,height=2.5in]{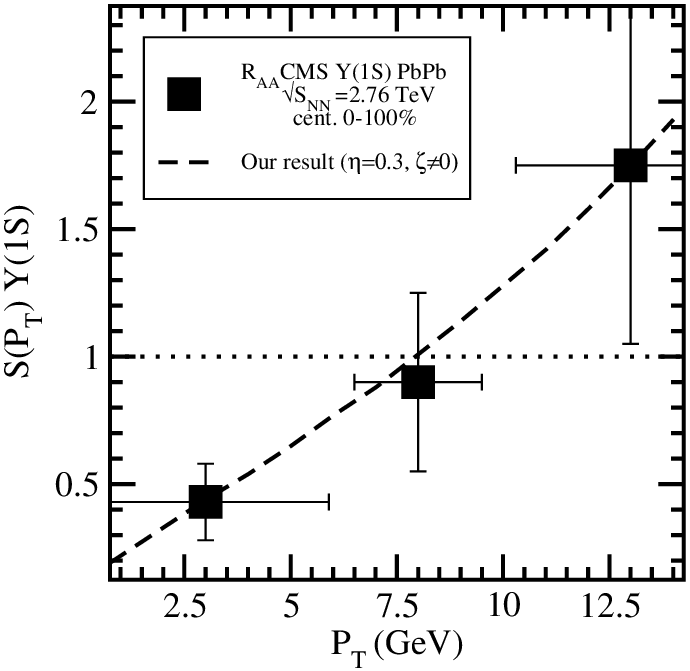} \\ 
\end{array}$
\end{center}
\caption{\footnotesize Centrality dependence of suppression factor for
$\Upsilon$(1S) (left panel) and variation of suppression factor against $P_T$
for $\Upsilon$(1S)(right penal).}
\end{figure}
To study the centrality dependence of the suppression factor, 
we use the CMS measurements of the pseudo rapidity and 
centrality dependent transverse energy density in Pb-Pb collisions
at the LHC energy~\cite{Phenix:Collab}, to obtain the initial
condition:
\be
\langle \epsilon_i \rangle=\frac{\xi}{A_T c \tau_i}J(y,\eta)\frac{dE_T}{d\eta}~,
\label{eqdet}
\ee 
where the Jacobian $J(y,\eta)$ (=1.09) is taken from HYDJET 1.8 for 
the pseudorapidity range $|{\eta}|< $ 0.35 in central collisions
at $\sqrt{S_{NN}}$= 2.76 TeV~\cite{CMS:Collab}. 

For the top 5\% central (Pb-Pb) collisions at the LHC, the Bjorken 
formula (\ref{eqdet}) (without the factor, $\xi$) estimates the (initial) 
energy density, $<\epsilon>_i$ = 14 GeV/$fm^3$ for initial time $\tau_i$=1 fm. 
Although this estimate is 2.6 times larger than that at RHIC 
energy~\cite{Phenix:Collab}, but it underestimates the initial energy 
density to the extent such that even the exited states of $\Upsilon$
family have not been dissolved by the deconfined medium. So a scale 
factor ($\xi \sim 5 $) has been introduced in the Bjorken formula to get
rid of the unusually smaller values~\cite{Hirano:PRC2001}.
\begin{figure}[h]
\vspace{1.55in}
\begin{center}$
\begin{array}{cc}
\includegraphics[width=2.5in,height=2.5in]{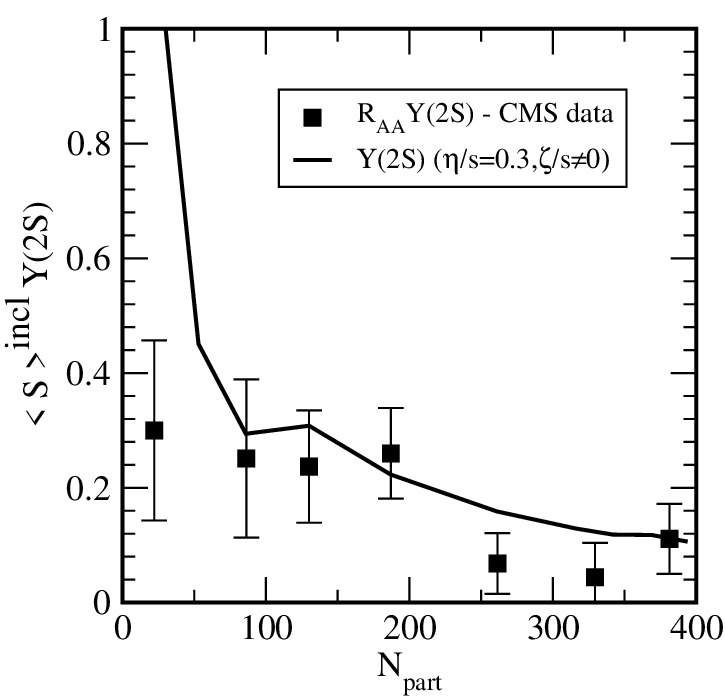} & \hspace{0.25in}
\includegraphics[width=2.5in,height=2.5in]{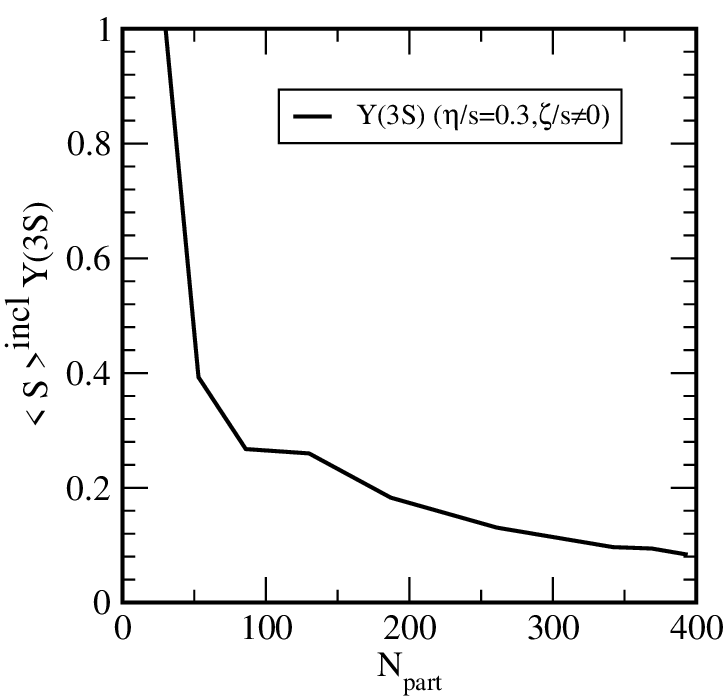}\\
\end{array}$
\end{center}
\caption{\footnotesize The centrality dependence of the inclusive survival
probability of the 2S(left panel) and 3S(right panel) states, respectively.}
\end{figure}
\begin{figure}[h]
\vspace{.1in}
\begin{center}$
\begin{array}{cc}
\includegraphics[width=2.5in,height=2.5in]{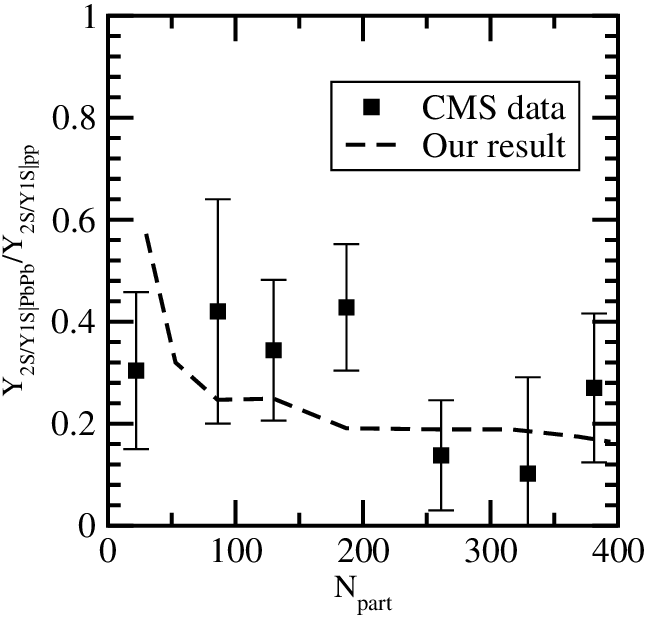} & \hspace{0.25in}
\includegraphics[width=2.5in,height=2.5in]{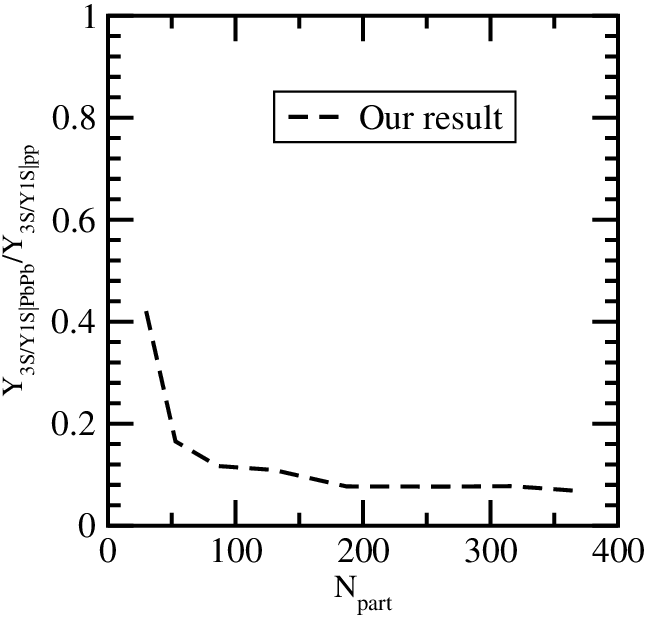} \\ 
\end{array}$
\end{center}
\caption{\footnotesize Centrality dependence of the double ratio  
$\frac{\Upsilon(2S)/\Upsilon(1S)|PbPb}{\Upsilon(2S)/\Upsilon(1S)|pp}$ 
(left panel) and
$\frac{\Upsilon(3S)/\Upsilon(1S)|PbPb}{\Upsilon(3S)/\Upsilon(1S)|pp}$
(right panel) respectively.}
\end{figure}
\begin{figure}[h]
\vspace{.1in}
\begin{center}$
\begin{array}{cc}
\includegraphics[width=2.5in,height=2.5in]{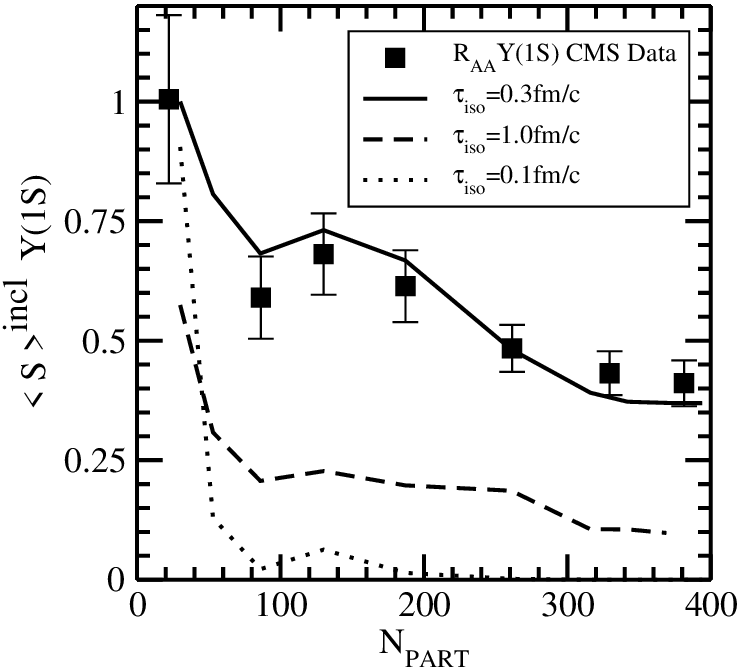}
\end{array}$
\end{center}
\caption{\footnotesize Centrality dependence of the sequential 
suppression of $\Upsilon$(1S) for different isotropization 
times}
\end{figure}
\subsection{Results and discussions}
Quarkonium suppression in nucleus-nucleus collisions compared to $p$-$p$ 
collisions involves various time-scales, associated with a) the initial
conditions of the medium, b) the dynamics of the expansion, and the in-medium 
properties of the quarkonium states and finally the competition among them 
ensues a rich structure in the suppression pattern. The first one is related
to the time scale of thermalization, the second one is associated to the
formation of resonances in the dilated (fireball) frame ($t_F$), which 
depend on 
their intrinsic transverse momenta and the formation times in their 
rest frame, and are related through a hierarchy: $\tau_F (1S)<\tau_F(2S)<\tau_F(3S)$.
The second is related to the expansion rate of the medium that can be
effectively gauged in terms of the speed of sound, which, in turn,
is interconnected through the equation of state ($p=c_s^2 \epsilon$).
The third one is the screening time, $\tau_s$ (the time-span
of the suppression) which depends on the scale of dissociation (the screening
energy, $\epsilon_s$) and the speed of sound, $c_s$.
The time $\tau_s$ also depends on the centrality of the collision (initial
conditions), i.e., if the
collision is more central then the system starts initially
from the higher energy density and take longer time
to reach $\epsilon_s$. On the other hand, since the excited states 
are dissociated at lower temperatures compared to the ground state,
so $\epsilon_s$'s satisfy the hierarchy:
$\epsilon_s (1S)>> \epsilon_s (2S) >> \epsilon_s (3S)$, hence the
screening times, $\tau_s$'s will thus satisfy the reverse relation: $\tau_s(1S)<\tau_s(2S)<\tau_s(3S)$.
However, the hierarchy in the screening times, in conjunction 
with the formation times makes
the suppression pattern complicated, {\em for example}
the suppression of  $\Upsilon$(2S) state may not always larger than 
 $\Upsilon$(1S) state and the $\Upsilon$(3S) state may not be suppressed 
 more than the $\Upsilon$(2S) state.

We will now discuss how the competition of the various time-scales 
transpires into the suppression pattern. Suppose if $\epsilon_s \gtrsim 
\epsilon_i$, there will be no suppression and if 
$\epsilon_i \gtrsim \epsilon_s$,
there will be suppression but the extent of suppression depends on 
i) how big the difference, $\Delta$ (=$\epsilon_i- \epsilon_s)$ is between
$\epsilon_i$ and $\epsilon_s$ and it varies from one state to other, 
and ii) how fast the system reaches to different $\epsilon_s$'s, i.e., how 
large the screening time, $\tau_s$ is for different states,
which in turn can be modulated by the  
bulk and shear forces near and away from the critical temperature, 
respectively. For a fixed centrality ($\epsilon_i$), $\Delta$ is 
minimum for $\Upsilon$ (1S) and increases for the excited 
states due to the hierarchy in $\epsilon_s$'s. For a fixed $\Delta$, 
$\tau_s$ becomes larger due to the presence of dissipative forces, compared
to ideal fluid. In fact, the shear viscosity slows down the expansion 
at the early stages of the expansion and thus affect the screening 
of the $\Upsilon$(1S) most whereas the bulk viscosity slows
down the late stages of the expansion and hence the excited states are 
affected much.
This is due to the fact that the shear viscosity is developed at the early 
stages and diminishes gradually whereas the bulk viscosity sets in
late in the proximity of the critical temperature.
Thus both the bulk and shear viscosities act as an additional handle to
decipher the suppression pattern.

With this understanding, we now analyze the results on the screening energy 
density contours for the $\Upsilon$ (1S) and $\Upsilon$ (2S) states
in Fig.3 at LHC energy, i.e. how the topology of the contour depends 
on the quarkonium properties, the expansion dynamics etc. 
The main observations are: i) the size of the 
contour increases while going from the ground state to the excited 
states because the screening energy density decreases from  $\Upsilon$(1S) 
to  $\Upsilon$(2S) states rapidly, so the system takes longer to reach 
$\epsilon_s$ for the excited states, ii) 
the contour also increases with the increase of the 
viscous forces because the viscous forces
slows down the entire evolution. More 
specifically the contour of the ground states are affected by the shear 
term only, whereas the excited states are affected by both.
The above observations can be encrypted in the boundary of the screening 
region ($r_s$). To understand both the centrality ($N_{\rm{Part}}$) and 
transverse momentum ($p_T$) spectrum of the suppression pattern, 
we have calculated $r_s$ as a function of the number
of participants ($N_{\rm{Part}}$) for various $p_T$'s in Fig. 4.
It is found that the size of the screening boundary ($r_s$) 
increases rapidly with the centrality and explains why there is more 
suppression in most central collision and less suppression in peripheral
collision. To be more specific, for 1S state, the screening boundary initially
enlarges rapidly and gets saturated for $N_{\rm{part}} \ge 300$, while 
for the excited states it increases monotonically, with the centrality. 
This explains why there is a saturation trend in the CMS results for the 
$\Upsilon$ (1S) suppression for $N_{\rm{part}}>300$ (left panel of Fig. 5)
and there is gradual suppression for the excited (2S and 3S) states 
(Fig. 6). We also notice that for a given 
centrality, the screening boundary for the $\Upsilon$ (1S) state gets 
squeezed rapidly for $Q \bar Q$ pairs having larger $p_T$, while the boundary 
gets swelled for pairs with smaller momenta. Thus for smaller number
of participants, the $p_T$ above which a pair can escape, is larger
than the larger number of participants.  
Since the production of partons with smaller $p_T$'s are more abundant
in smaller centralities than the higher centralities, so the above
observation explains why there is  more suppression even 
in the smaller centralities (left panel of Fig. 5). 
However, the sensitivity of the transverse momenta is less prominent 
for the excited states (right panel of Fig. 4). 
That is why there is no strong centrality dependence 
in the suppression pattern for $\Upsilon$ (2S) and (3S) states (Fig 6), 
in contrast to the $\Upsilon$(1S) state.

With these ingredients, we explain our results on the inclusive 
survival probability for the $\Upsilon$ (1S) state 
computed from the feed down of the exited states (left panel of Fig. 5).
We found that the suppression increases with the centrality upto
$N_{\rm Part}$ =300 and nearly saturates beyond $N_{part} >350$.
This finding is compatible with our earlier observation 
(left panel of Fig. 4), where the screening radius ($r_s$)
is almost independent of the centrality beyond a certain value. 
We also notice that the inclusive (survival) probability (averaged over
the centralities) for $\Upsilon$(1S) state increases linearly with $p_T$
(right panel of Fig. 5). This is again compatible with 
our earlier observation, where for a given centrality, $r_s$ 
increases almost linearly with $p_T$.

We have also calculated the inclusive survival probability 
for the $\Upsilon$ (2S) and (3S) states (Fig. 6), which 
are found to decrease slowly with the centrality.
This finding resonates with the earlier observation 
(right panel of Fig. 4), where for a given $p_T$, the screening radius ($r_s$)
increases linearly with the centrality and for a given centrality, the
$p_T$ dependence of $r_s$ is very slow. 

The initial state effects affect the $\Upsilon$ (nS) states in a similar 
manner, so the possible acceptance and/or efficiency differences 
cancel out in the ratio, ${\Upsilon(nS)/\Upsilon(1S)}_{PbPb}$
(with respect to the $p$-$p$ collisions). Moreover the final state nuclear 
absorption effects are expected to minimum at LHC 
energies~\cite{zwlin:PLB2001}, so we have calculated the double 
ratio (Fig. 7) at the LHC energy, which shows poor suppression of 
2S state with respect to 1S state for peripheral collision and 
no characteristic dependence on the collision centrality for 
$N_{Part}>100$. But CMS results
show more suppression of $\Upsilon$(2S) for peripheral collisions and other 
approaches ~\cite{Tsong:PRC852012,Aemerick:EPJA482012,
Fnendzig:arxiv12108366v2} also agrees
with this fact. This indicate that either their may be some additional 
suppression mechanisms  which are still missing  in the theoretical
 calculations or CMS  measurements are not sufficient to disentangle
the nuclear effects from medium effects and it could be 
better resolved on availability  of more data from  heavy-ion 
and proton-nucleus collision runs at LHC in future. 
The results of double ratio shows exited 
states are suppressed more with respect to ground state.

To explore the effects of the viscous forces on the suppression, we take 
the shear viscosity to entropy ratio, $\eta/s$ as 0.08 and 0.3 along with 
the parametrization of the bulk viscosity, $\zeta/s$ from (\ref{eq:zetas}).
We notice that the suppression increases with the increase of 
shear viscosity which, in turn, enhances the screening time.
Our estimate agrees with the CMS data when the ratio $\eta/s$ is
taken from its perturbative estimate. This seems justified because 
the screening energy density for the $\Upsilon$(1S) state is very high
where the perturbative calculation seems meaningful. So the $\Upsilon$ (1S) 
production can be used to constrain the $\eta/s$ ratio.

Since the physics of isotropization is yet to understand theoretically, so
the duration of the pre-equilibrium era is uncertain.
Therefore we take the privilege to constrain the arbitrariness of
$\tau_{\rm{iso}}$ by the suppression of bottomonium production 
( Fig. 8) because $\Upsilon$ (1S) is formed earlier
than the isotropization time, which is not the case for $\Upsilon$ (2S)
state. We have found that $\tau_{iso}$ =0.3 fm looks 
more plausible as far as CMS data is concerned.

\section{Conclusions}
In conclusion we have studied the sequential suppression for $\Upsilon$ (1S)
and $\Upsilon$ (2S) states at the LHC energy in a longitudinally expanding 
partonic system, which underwent through the successive 
pre-equilibrium and equilibrium phases in the presence of dissipative forces.
Quarkonium suppression in nucleus-nucleus collisions
compared to $p$-$p$ collisions couples the
in-medium properties of the quarkonium states with 
the dynamics of the expanding medium.
In this work we obtained the dissociation temperatures of the quarkonium 
states by correcting both 
the perturbative and nonperturbative terms in $Q\bar Q$ potential 
in (an)isotropic medium through the HTL resummed perturbation.
We then modeled the pre-equilibrium evolution as anisotropic fluid 
via the time dependent anisotropic parameter, 
$\xi(\tau)$ and hard momentum scale, $P_{ hard}(\tau)$ while
the equilibrium era is governed by the second-order
dissipative hydrodynamics in (1+1) Bjorken boost-invariant 
model and coupled them together to estimate the sequential suppression.
The expansion in equilibrium hydrodynamics is controlled by the speed 
of sound $c_{s}^{2}$, which could be further handled by the lattice 
QCD equation of state, the shear ($\eta$) and bulk ($\zeta$) viscous 
forces etc.

The bulk viscosity in conjunction with the
shear viscosity  enhances the cooling rate and thus causes
more suppression to the ground state, however, the bulk viscosity
$\zeta/s$ is significant for exited states.
The sequential suppression  is a very complex 
phenomenon depends on several parameters 
such as the scale of the dissociation of quarkonium states, the decay
of the excited states, the centrality of collision, the transverse 
momentum, the screening time, the formation time, the dissipative forces
etc., including the isotropization time and too early or too late 
isotropization results in over suppression. 
The tiny formation time (compared to the isotropization time)
and tightly bound character of the bottomonium states help the
the suppression of bottomonium to constrain
both the isotropization time (0.3 fm) as well as the shear 
viscosity-to-the entropy ratio (0.3)

\noindent {\bf Acknowledgments:}
One of us (BKP) is thankful for some financial assistance 
from CSIR project (CSR-656-PHY), Government of India. USK is also
thankful to Government of Maharashtra for the financial assistance.

\end{document}